\documentclass[preprint,aps,prb,showpacs,preprintnumbers,amsmath,amssymb]{revtex4-1}

\usepackage{graphics}
\usepackage{epsfig}
\usepackage{bm}

\begin{document}

\title{Mechanocaloric effects in Shape Memory Alloys.}

\author{Llu\'{\i}s Ma\~nosa and Antoni Planes}

\affiliation{Departament d'Estructura i Constituents de la Mat\`eria. Facultat de F\'{\i}sica. Universitat de Barcelona. Mart\'{\i} i Franqu\`es, n. 1. 08028 Barcelona. Catalonia.\\}




\begin{abstract}
Shape memory alloys are a class of ferroic materials which undergo a structural (martensitic) transition where the associated ferroic property is a lattice distortion (strain). The sensitiveness of the transition to the conjugated external field (stress), together with the latent heat of the transition gives rise to giant mechanocaloric effects. In non-magnetic shape memory alloys, the lattice distortion is mostly described by a pure shear and the martensitic transition in this family of alloys is strongly affected by uniaxial stress whereas it is basically insensitive to hydrostatic pressure. As a result, non-magnetic alloys exhibit giant elastocaloric effects but negligible barocaloric effects. By contrast, in a number of magnetic shape memory alloys, the lattice distortion at the martensitic transition involves a volume change in addition to the shear strain. Those alloys are affected by both uniaxial stress and hydrostatic pressure and they exhibit giant elastocaloric and barocaloric effects.  The paper aims at providing a critical survey of available experimental data on elastocaloric and  barocaloric effects in magnetic and non-magnetic shape memory alloys.

\end{abstract}

\maketitle

\section{Introduction}

Mechanocaloric effects refer to the thermal response (adiabatic temperature  and isothermal entropy changes) of a solid when subjected to an external stress. When the solid is in the vicinity of a structural first-order transition, these quantities can reach large values for moderate stresses giving rise to the so-called giant caloric effects \cite{Fahler2012,Manosa2013a,Moya2014,Moya2015}. These large values are related to the release (or absorption) of the latent heat associated with the first-order transition. From an applied point of view, giant caloric materials are good candidates for the development of an efficient environmental friendly solid state refrigeration technology \cite{Kitanovski2015}. 

Materials exhibiting giant mechanocaloric effects add up to the family of materials with giant magnetocaloric and electrocaloric effects. While magnetic field and magnetization, as well as electric field and polarization, have a vectorial character (rank-one tensors), stress and strain are rank-two tensors. This means that even for isotropic materials a complete characterization of the mechanocaloric effect requires determination of the solid response to two independent stresses, in contrast to magnetocaloric and electrocaloric effects which are fully characterized by the response to the magnetic or electric field applied on a given direction. Experimentally, the typical stresses applied to measure mechanocaloric effects are  uniaxial load and hydrostatic pressure. The thermal response to uniaxial load is known as elastocaloric effect and to a hydrostatic pressure, barocaloric effect.

Prominent among the materials exhibiting giant mechanocaloric effects are shape memory alloys. On cooling, these alloys undergo a transition from a high-temperature (high symmetry) cubic structure towards a lower-temperature (lower symmetry) close-packed structure. This is the martensitic transition which is first-order and difusionless, with a lattice distortion which is mostly given by a shear. Such a shear distortion is significantly large and makes the martensitic transition strongly sensitive to application of external uniaxial stress, which gives rise to a series of unique thermomechanical properties exhibited by these alloys such as shape memory effect, pseudoelasticity and superelasticity \cite{Otsuka2002}. More recently \cite{Bonnot2008} it has also been shown that shape memory alloys do also exhibit giant elastocaloric effects. 

Shape memory alloys can be broadly classed into two categories: conventional (non-magnetic) and magnetic. While in non-magnetic alloys the volume of the cubic unit cell is very close to that of the martensitic unit cell, in magnetic alloys the coupling between magnetic and structural degrees of freedom can give rise to  relatively large volume changes (which can reach $\sim $ 0.5-1 \% ) at the martensitic transition. The martensitic transition in conventional shape memory alloys is therefore largely insensitive to hydrostatic pressure and these compounds do not present significant barocaloric effects. However in magnetic shape memory alloys the transition can by affected by hydrostatic pressure \cite{Manosa2008} and these alloys are prone to exhibit barocaloric effects \cite{Manosa2010}.

Another interesting peculiarity of magnetic shape memory alloys is the interplay between structural and magnetic degrees of freedom which enables the transition to be driven by the application of any of the fields conjugated to the order parameters of the transition (uniaxial stress, hydrostatic pressure and magnetic field). This property gives rise to the so-called multicaloric effects \cite{Moya2014,Planes2014} which offer new possibilities in the prospects for solid-state refrigeration.  In the present paper we will overview  elastocaloric effects associated with the application of uniaxial stress in conventional shape memory alloys, and both elastocaloric and barocaloric effects in magnetic shape memory alloys.

\section{The measurement of mechanocaloric effects}

The isothermal entropy change ($\Delta S$) and the adiabatic temperature change ($\Delta T$) associated with the application (or removal) of an external stress can be derived from experiments by a variety of measurement protocols which are broadly classed into indirect, quasi-direct and direct methods. 

Indirect methods involve the measurement of the temperature and stress dependence of the strain and rely on the use of the Maxwell relations.  Isofield temperature scans that cover the whole temperature range over which the transition spreads are preferable than isothermal stress scans. This is because the former are free from a possible overestimation of entropy changes which may occur when isothermal measurements are carried out in a mixed two-phase state. For elastocaloric materials, length changes across the transition can be measured by suitable strain gauges and the entropy change (per mass unit) can then be computed as:

\begin{equation}
\Delta S(T,F) = \frac{1}{m} \int_{0}^{F} \left( \frac{\partial L}{\partial T} \right)_{F} dF
\label{SindirectL}
\end{equation}

\noindent where $F$ is the applied uniaxial load, and $m$ and $L$ are, respectively, the mass and the gauge length of the specimen. An alternative expression in terms of strain $\epsilon$ and stress $\sigma$ is:

\begin{equation}
\Delta S(T,\sigma) = \frac{1}{\rho} \int_{0}^{\sigma} \left( \frac{\partial \epsilon}{\partial T} \right)_{\sigma} d\sigma
\label{Sindirectepsilon}
\end{equation}

\noindent where $\rho$ is the mass density, $\sigma \simeq F/A$ (with $A$ the cross section which is assumed to be constant) and $\epsilon = (L-L_0)/L_0$, with $L_0$, the gauge length at zero stress. In the case of the barocaloric effect, volume changes across structural transitions are typically small and it becomes difficult to perform suitable measurements with enough accuracy to allow a reliable numerical computation of $\Delta S$ using equations \ref{SindirectL} and \ref{Sindirectepsilon}.

Quasi-direct methods are based on differential scanning calorimety (DSC) under applied external fields. In that case, DSC heating and cooling runs are performed at different (constant) values of the applied external stress. This kind of calorimeters are available in the case of barocaloric effects (DSC under hydrostatic pressure) but, at present, no suitable DSCs are available that can operate under uniaxial external loads. Data from DSC under field are complemented with specific heat ($C$) data away from the structural transition and it is assumed that in those regions $C$ is not significantly influenced by pressure. From these data, the entropy $S(T,p)$ (referenced to the value at a given temperature $T_0$)  is computed. Then, the entropy change associated with the application of a pressure $p$ is obtained by subtracting the  $S(T,p)$ curves computed at different values of pressure:

\begin{equation}
\Delta S(T,0 \rightarrow p) = S(T,p) - S(T,0)
\end{equation}

\noindent (where $p=0$ refers to atmospheric pressure). 

This method also permits computing the adiabatic temperature change by subtracting the corresponding $T(S,p)$ curves:

\begin{equation}
\Delta T(S,0 \rightarrow p) = T(S,p) - T(S,0)
\end{equation}

Direct determination of entropy changes require the use of DSCs  under field that can operate isothermally while the field is scanned \cite{Casanova2005,SternTaulats2014a,Moya2013}. Unfortunately until now this technique is not available for the study of mechanocaloric effects. On the other hand, direct measurements of adiabatic temperature changes can be performed by suitable thermometers attached to the studied sample or alternatively by means of non-contact infrared thermometry. The adiabaticity of these measurements relies on the ratio between the characteristic time constant associated with the application (or removal) of the stress and the time constant associated with the heat exchange between sample and surroundings. In the case of barocaloric effects, the sample is surrounded by the pressure-transmitting fluid and measurements are not fully adiabatic thereby leading to measured $\Delta T$ values which are always underestimated. However, in most elastocaloric experiments the sample is typically in air, and application (or removal) of uniaxial stresses at strain rates greater than 0.1 s$^{-1}$ have proved to be close to the adiabatic limit and therefore provide reliable data for the adiabatic temperature change.

It is also worth mentioning that it is customary to estimate temperature changes from measured entropy changes (or alternatively entropy changes from measured temperature data) by means of the relationship:

\begin{equation}
\Delta T \simeq - \frac{T}{C} \Delta S
\label{estimated}
\end{equation}

\noindent It must be taken into account, however, that such an estimation may provide data which are overestimated for $\Delta T$ and underestimated for $\Delta S$ \cite{Sandeman2012}.


\section{Conventional (non-magnetic) shape memory alloys.}

There are two major families of non-magnetic shape memory alloys: Cu-based and Ti-Ni based alloys. The lattice distortion at the martensitic transition corresponds to a pure shear of the $\{110\}$ planes along the $<1\bar{1}0>$ directions and the volume change is negligible small \cite{Otsuka1998}. 
The open bcc structure has low-energy transverse TA$_2$ phonons which confer to this phase a large vibrational entropy \cite{Planes2001} whereas the vibrational entropy of the martensitic phase is lower. Hence the major contribution to the transition entropy change giving rise to the giant mechanocaloric effect in these alloys has a vibrational (phonon) origin.   

Cu-based alloys transform from a DO$_3$ or L2$_1$ ($Fm3m$) ordered structures to a monoclinic one whose structure can alternatively be described by a larger unit cell which can be approximated to be orthorhombic (18R, $I2/m$) or hexagonal (2H, $Pnmm$) for which the monoclinic angle is  very small \cite{Ahlers1986}. Depending on the alloy composition, the martensitic phase can exhibit a variety of structures which differ in the modulation of the close-packed planes.

\begin{figure}[!h]
\centering\includegraphics[width=12cm]{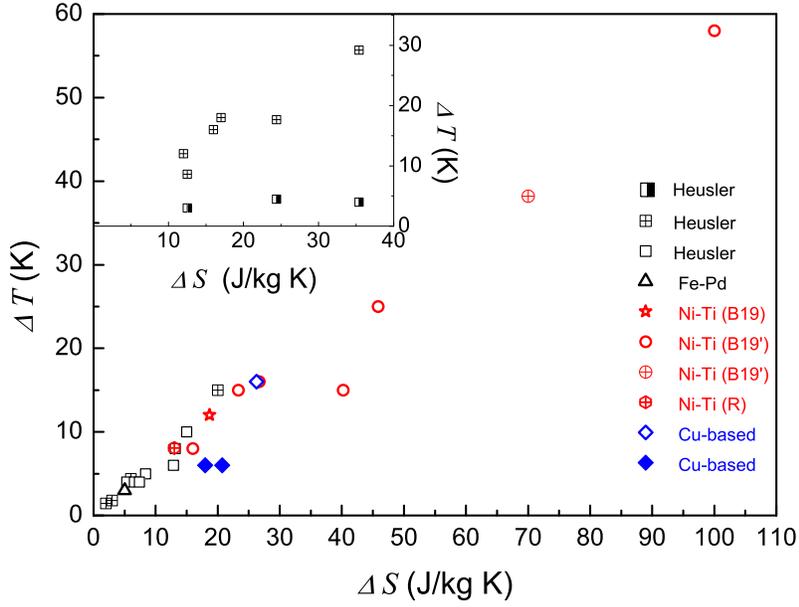}
\caption{Adiabatic temperature change as a function of the isothermal entropy change associated with the elastocaloric effect for a variety of shape memory alloys. The inset corresponds to the values associated with the barocaloric effect. Open symbols correspond to directly measured $\Delta T$, and estimated $\Delta S$ via equation \ref{estimated}. Crossed symbols correspond to calorimetrically measured $\Delta S$, and estimated $\Delta T$ via equation \ref{estimated}. Half filled symbols correspond to $\Delta S$ and $\Delta T$ derived from calorimetric measurements. Solid symbols correspond to directly measured $\Delta T$ and indirectly measured $\Delta S$.}
\label{figure1}
\end{figure}

First results of a giant elastocaloric effect in shape memory alloys were reported on a Cu-Zn-Al single crystal \cite{Bonnot2008} where isothermal entropy changes were computed from stress-strain experiments. Later experiments \cite{Vives2011,Manosa2009,Manosa2013b,Tusek2015a} have been performed on both single crystals and polycrystals in tensile and compressive modes, including direct measurements of adiabatic temperature changes. As a summary, those experiments have proved that Cu-based alloys exhibit a large isothermal entropy change ($\Delta S \sim $ 20 J/kg K) which can be achieved already at very low values of the applied stress (around 25 MPa for experiments carried out at temperatures close to the stress-free martensitic transition temperature). Such an entropy value coincides with the transition entropy change ($\Delta S_t$), which is very weakly dependent upon composition \cite{Obrado1997}, and represents the upper bound for the elastocaloric effect in these compounds. With regards to the temperature change, some experiments lack perfect adiabaticity because the grips act as thermal sinks and the measured values (solid blue diamonds in fig. \ref{figure1}) are lower than expected. However, for long enough samples good adiabaticity can be achieved and reported values ($\Delta T \sim $10-14 K, open blue diamond in fig. \ref{figure1}) \cite{Rodriguez1980,Vivesunp} approach the maximum predicted value given by equation \ref{estimated}, taking $\Delta S_t$ as the uppper bound.  

\begin{figure}[!h]
\includegraphics[width=12cm]{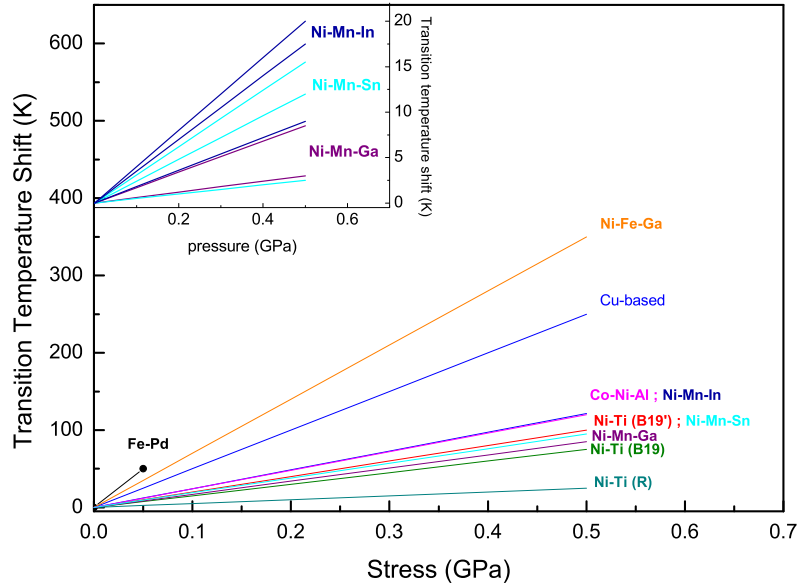}
\caption{Shift in the transition temperature with the applied uniaxial stress for illustrative shape memory alloys. The inset shows the shift in the transition temperature under hydrostatic pressure.}
\label{figure2}
\end{figure}

\begin{figure}[!h]
\includegraphics[width=12cm]{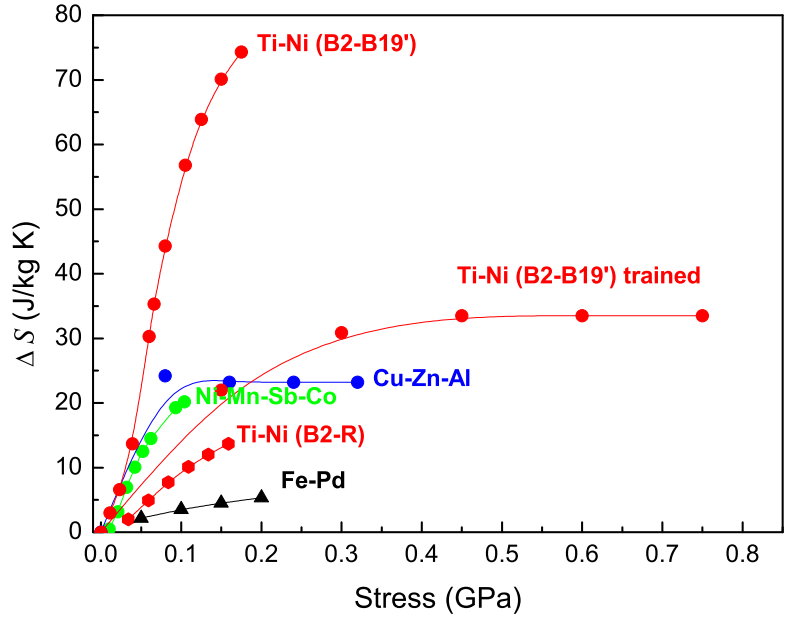}
\caption{Maximum value for the stress-induced isothermal entropy change as a function of uniaxial stress for the elastocaloric effect in a variety of shape memory alloys (as labelled in the figure).}
\label{figure3}
\end{figure}

Another important family of conventional SMA is the one based on Ti-Ni. These alloys transform from a B2 ($Pm3m$) ordered cubic structure to a monoclinic B19' ($P2/m$) martensitic phase. It must be taken into account, however, that depending on the specific heat treatment and doping with a third element, the transformation path can take place via intermediate structural phases. For suitably annealed samples, the cubic structure transforms towards a trigonal R ($P3$) phase, which upon further cooling transforms towards the monoclinic B19' phase. The reverse transformation upon heating takes place in a single step from B19' to B2. On the other hand, in Cu-doped samples (with Cu$\geq$ 5\%), the B2 phase transforms towards an orthorhombic B19 ($Pmma$) martensitic phase. A thorough description of the details of the martensitic transition in Ti-Ni alloys is beyond the scope of the present paper, and they can be found in ref. \cite{Otsuka2005}. For present purposes it is important to recall that B2 $\leftrightarrow$ B19' transition has a large latent heat (with a transition entropy change $\Delta S_t $ in the range 60-90 J/kg K), and a large hysteresis \cite{Frenzel2015,SotoParra2016}. The B2 $\leftrightarrow$ B19 has a slightly lower latent heat ($\Delta S_t \sim$ 50-60 J/kg K), and a lower hysteresis \cite{Frenzel2015}. The B2 $\leftrightarrow$ R transition has a much lower latent heat ($\Delta S_t \sim$ 15-20 J/kg K) and a very reduced hysteresis \cite{SotoParra2016}. It is also worth pointing out that B2 $\leftrightarrow$ R and B2 $\leftrightarrow$ B19 transitions exhibit an excellent reproducibility upon thermal and stress cycling. The critical stress to induce the B2 $\rightarrow$ B19' transition is larger than for the  B2 $\rightarrow$ B19 and B2 $\rightarrow$ R transitions, which means that typically stresses above 500 MPa are required to achieve a giant elastocaloric effec associated with the B2 $\leftrightarrow$ B19' transition,  whereas the B $\rightarrow$ R  transition requires  much lower stresses.

The elastocaloric properties of Ti-Ni-based shape memory alloys have been studied by several research groups \cite{Tusek2015a,Cui2012,Pieczyska2013,Tusek2015b,SotoParra2016,Ossmer2014,Ossmer2015a,Ossmer2015b,Schmidt2015,
Bechtold2012,Pataky2015,Tang2015}. Experiments have been carried out in tensile and compressive modes in single crystals and polycrystals and different shapes such as wires, films, etc have been the subject of investigation. Most of the studies correspond to direct adiabatic temperature measurements and only a few of them report isothermal entropy data. The behaviour exhibited by films and wires compares well with that exhibited by bulk samples (including single crystals). The B2 $\leftrightarrow$ B19' transition provides the largest values for $\Delta T$ (red circles in fig. \ref{figure1}), with measured values that can reach 25-60 K \cite{Bechtold2012,Cui2012,Pieczyska2013,Tusek2015a,Tusek2015b,Ossmer2015b}. Values for the B2 $\leftrightarrow$ B19 are lower (red star in fig. \ref{figure1}), in the range 10-15 K \cite{Bechtold2012,Ossmer2015a}. No direct $\Delta T$ measurements are reported for the B2 $\leftrightarrow$ R transition, but based on the entropy data \cite{SotoParra2016}, values are expected to be around 10 K (red hexagon in fig. \ref{figure1}). With regards to the entropy change, large values ($\Delta S \sim $ 60-80 J/kg K)  have been reported for the B2 $\leftrightarrow$ B19' transition although $\Delta S$ values can be reduced in samples that have been trained for many cycles \cite{Tusek2015b}. The isothermal entropy change associated with the B2 $\leftrightarrow$ R transition is $\Delta S \sim$ 13 J/kg K. It is worth noting that although the B2 $\leftrightarrow$ B19' transition has a much larger entropy change, this value is not reproducible when the sample is cycled at low stress values, while an excellent reproducibility is obtained for the B2 $\leftrightarrow$ R transition \cite{SotoParra2016}. No entropy data are available for the B2 $\leftrightarrow$ B19 transition, but the maximum (which is bounded by the latent heat of the transition) is expected to be $\sim$ 50-60 J/kg K.

A comparison of the properties of Cu-based and Ti-Ni based alloys can be found in Figures \ref{figure2} and \ref{figure3}. Figure \ref{figure2} shows that the transition temperature in Cu-based alloys  is more sensitive to the applied stress than in Ni-Ti alloys. The shift in the transition temperature with stress is given by the Clausius-Clapeyron equation $dT/d\sigma = - v \Delta \epsilon / \Delta S_t$ where $v$ is the specific volume, 
$\Delta \epsilon$, the transition strain (determined by the crystallographic change) and 
$\Delta S_t$, the transition entropy change. While $\Delta \epsilon $ for Cu-based alloys is comparable to that 
for the  B2 $\leftrightarrow$ B19' transition ($\Delta \epsilon \sim $ 7-10 \%), the lower $\Delta S_t$ for Cu-based alloys results in a larger $dT/d\sigma$.  With regards to the  B2 $\leftrightarrow$ R transition the low $dT / d\sigma$ values are due to a reduced transition shear strain ($\Delta \epsilon \sim$ 0.5-1 \%). 

Figure \ref{figure3} compares the elastocaloric effect for Cu-based and Ti-Ni-based alloys in terms of the maximum value for the isothermal stress-induced entropy change, as a function of stress.  Data have been obtained using the indirect method (stress vs. temperature curves at fixed values of stress). The B2 $\leftrightarrow$ B19' transition in Ti-Ni exhibits large values which are significantly reduced for trained samples. It is also worth taking into account that these large values are not reproducible under stress cycling for low values of the stress (below 500 MPa). A distinctive feature for Cu-based alloys is that the maximum $\Delta S$ is achieved for very low stress values  ($\sigma \leq $ 100 MPa). These values are reproducible under stress cycling for stresses as low as 100 MPa. The B2 $\leftrightarrow$ R exhibits moderate values, which for stresses up to 200 MPa are below the value corresponding to the transition entropy change, indicating that for these values of the stress it is not possible to achieve transformation of the whole sample.


\section{Magnetic shape memory alloys.}

The martensitic transition in Fe-based magnetic  alloys was studied many decades ago and some of these alloys, like ordered Fe-Pt and Fe-Pd were shown to exhibit shape memory properties \cite{Kakeshita2002}. However, it was with the report of magnetic shape memory in a Ni-Mn-Ga single crystal \cite{Ullakko1996} that the research in magnetic shape memory alloys experienced an enormous increase. Since then, a plethora of Ni-based (and some Co-based) Heusler alloys have been reported to exhibit martensitic transitions with associated magnetic shape memory \cite{Planes2009,Acet2011}. For magnetic shape memory alloys (MSMA) the high temperature cubic phase is ferromagnetic, and depending on composition the martensitic transition can take place above or below the Curie point. The interest is mostly focussed on those alloys transforming martensitically close or below the Curie point because they exhibit a variety of interesting properties related to the application of magnetic field. Depending on the magnetic order of the martensitic phase, magnetic shape memory alloys are broadly classed into two categories: alloys with a ferromagnetically ordered martensitic phase, and  alloys for which the martensitic phase is weakly magnetic (which are also known as  metamagnetic shape memory alloys). For metamagnetic shape memory alloys, the magnetic structure of the martensite is still a subject of debate \cite{Khovaylo2009,Ollefs2015} but the presence of short range antiferromagnetic interactions has been experimentally confirmed \cite{Aksoy2009}. In these metamagnetic shape memory alloys there is a large change in magnetization at the martensitic transition and therefore the transition temperature is strongly sensitive to magnetic field which gives rise to the magnetic superelasticity \cite{Kainuma2006,Krenke2007} and inverse magnetocaloric effects \cite{Krenke2005}.

From a crystallographic point of view, the high temperature phase is cubic ($Fm3m$). The martensitic structure in Fe-based alloys is a fct ($P4/mmm$) \cite{Cui2004}, and for Heusler compounds the martensitic structure can be tetragonal or modulated monoclinic, depending on alloy composition \cite{Acet2011}. Similar to conventional (non-magnetic alloys), the lattice distortion can be mostly described by a   shear of the $\{110\}$ planes along the $<1\bar{1}0>$ directions. However, in this case, the interplay between magnetism and structure can give rise to a volume change of the unit cell. Such a volume change is particularly relevant for those alloys with a large change in magnetization at the martensitic transition \cite{Aksoy2007} and therefore the martensitic transition in metamagnetic shape memory alloys will be sensitive to applied hydrostatic pressure \cite{Manosa2008} thereby giving rise to barocaloric effects, which will be discussed in the following sections.

In magnetic shape memory alloys there are two major contributions to the entropy: vibrational and magnetic. The leading contribution is the vibrational one, which, as occurs in non-magnetic alloys, has its origin on the low-energy TA$_2$ phonons which confer the high-temperature cubic phase a large vibrational entropy. For those alloys transforming between a ferromagnetic cubic phase and a ferromagnetic martensite the magnetic entropy plays a minor role. However, in metamagnetic shape memory alloys, the martensitic phase has a magnetic entropy larger than that of the cubic phase. Therefore the vibrational and magnetic degrees of freedom contribute in an opposite way to the transition entropy change. A detailed discussion on the significance of each contribution to the relative stability of cubic and martensitic phases can be found in ref. \cite{Kihara2014}.

\subsection{Elastocaloric effects.}

Reports on the elastocaloric properties of magnetic shape memory alloys are very recent \cite{Pataky2015,SotoParra2010,CastilloVilla2011,Xu2015,Xiao2015b,Huang2015,Lu2014,Lu2015,MillanSolsona2014,Sun2016,
CastilloVilla2013,Xiao2013,Xiao2015a}. 

Intermetallic magnetic Heusler alloys are typically very brittle and they cannot support stresses as large as those typically applied to non-magnetic shape memory alloys. Indeed, the first reports on Ni-Mn-Ga alloys (doped with Co and Fe to improve their toughness) were limited to very low stresses ($\sim$ 10 MPa), with a reduced stress-induced entropy change ($\sim$ 3-6 J/kg K) \cite{SotoParra2010,CastilloVilla2011}. Later, it was possible to develop tougher alloys by suitable doping and/or by preparing them as textured material or even single crystals, and stresses in the range 100-300 MPa have been investigated. Both magnetic shape memory and metamagnetic alloys have been the subject of investigation, and the majority of the results deal with direct measurements of adiabatic temperature changes with only a few of them reporting on entropy measurements. Illustrative data are compiled in figures \ref{figure1}, \ref{figure2} and \ref{figure3} where they are compared to data for the conventional SMA.

In most cases, the sensitivity of the transition temperature to uniaxial stress is lower than in Cu-based alloys and it is comparable to that of Ni-Ti (see fig. \ref{figure2}), with the exception of Ni-Fe-Ga single crystals stressed along the [001] direction. In that case, the large $dT/d\sigma$ arises form a very large transition strain $\epsilon \sim$ 14.5 \% and a moderate transition entropy change ($\Delta S_t \sim$ 16-20 J/kgK) \cite{Pataky2015}. Reported $\Delta T$ values (black squares in fig. \ref{figure1}) are in the range 2-15 K, lower than those corresponding to Cu-based and Ti-Ni-based alloys. Those lower values are likely to be due to the reduced values of the applied stress together with a moderate $dT/d\sigma$ which result in a partial transformation of the sample.  It is worth noting that the transition entropy change in many of these alloys may exceed that of Cu-based alloys with values that for suitable compositions can reach up to  $\Delta S_t \sim $ 80 J/kg K \cite{Krenke2005b,Krenke2006}.  Hence much larger $\Delta T$ could be expected if the applied stress was large enough to induce the transformation of the whole sample. With regards to the isothermal entropy change, in addition to the previously mentioned values at low stresses, data have only been reported for a Ni-Mn-Sb-Co alloy \cite{MillanSolsona2014}. $\Delta S$ data are shown as a function of the applied stress in figure \ref{figure3} in comparison with those for conventional shape memory alloys. A maximum value of $\Delta S  \sim$ 20 J/kg K was found for the maximum applied stress of 100 MPa. This value is about 60 \% of the total entropy change of that alloy (which provides the upper bound for the elastocaloric effect).

The behaviour for Fe-Pd is significantly different from that of the Heusler alloys \cite{Xiao2013,Xiao2015a}. The martensitic transition for this alloy is weakly first order (also termed second order like), with a low entropy change ($\Delta S_t \sim$ 1.2 J/kg K) and a reduced hysteresis. Nevertheless, the second-order characteristics of the transition imply that $\frac{\partial \epsilon}{\partial T}$ is not only given by the transition strain but it has also significant contributions beyond the transition region, and the resulting entropy (and temperature) change associated with the elastocaloric effect (see eqs. \ref{SindirectL} and \ref{Sindirectepsilon}) is not bounded by the latent heat of the transition.  The low $\Delta S_t$ combined with a moderate transition strain $\epsilon \sim$ 1.5 \% provides a significantly large $dT/d\sigma$ (see fig. \ref{figure2}), although that value must be taken cautiously because $\epsilon$ drastically changes with temperature and a critical point is expected at $\sigma_c \sim$ 40 MPa and $T_c \sim$ 280 K \cite{Xiao2015c}. The values for the elastocaloric effect (black triangle in fig. \ref{figure1}) are moderate (figs. \ref{figure1} and \ref{figure3}) $\Delta T \sim$ 3 K and $\Delta S \sim$ 4-5 J/kg K, but with an excellent reproducibility upon stress cycling.

\subsection{Barocaloric effects.}  
  


At present there are only a few reports on the barocaloric effect in magnetic shape memory alloys and until now data have been reported for Ni-Mn-In \cite{Manosa2010,SternTaulats2015a} and Ni-Co-Mn-Ga \cite{Manosa2014}. A reason for this scarcity of data is the need for purpose-built experimental systems \cite{Manosa2011} which are not usually at hand in many laboratories. Another  reason is that  among magnetic shape memory alloys only a few of them do have a volume change at the martensitic transition large enough to give rise to a significant barocaloric effect.  The strength of the barocaloric effect depends on the sensitivity of the martensitic transition to hydrostatic pressure, and as shown in the inset of figure \ref{figure2}, values for $dT/dp$ are in the range 5-40 K/GPa \cite{Manosa2008,Manosa2010,SternTaulats2015a,Albertini2007,Mandal2009,Kamarad2005,Yasuda2007,Muthu2011,
Sharma2011} which are about one order of magnitude smaller than the  $dT/d\sigma$ corresponding to uniaxial stress. Such a difference is a reflect that the major contribution to the lattice distortion at the martensitic transition is a shear strain.

\begin{figure}[!h]
\centering\includegraphics[width=12cm]{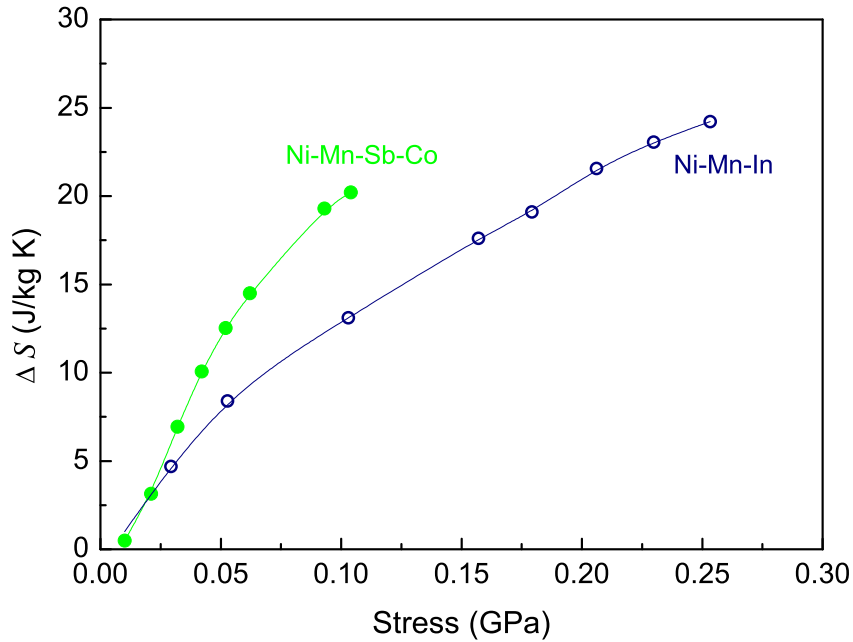}
\caption{A comparative between the maximum isothermal stress-induced entropy change for the elastocaloric and barocaloric effects, as a function of the applied stress. Solid green symbols stand for the elastocaloric effect in Ni-Mn-Sb-Co and open blue symbols, for the barocaloric effect in Ni-Mn-In.}
\label{figure4}
\end{figure}

Figure \ref{figure4} compares the barocaloric $\Delta S$ values in Ni-Mn-In to the corresponding data for the elastocaloric effect in Ni-Mn-Sb-Co. Although the barocaloric strength is lower, the final $\Delta S$ values are larger $\Delta S \sim$ 25 J/kg K because larger stresses can be applied in the hydrostatic case whereas the intrinsic brittelness of magnetic shape memory alloys limits the maximum uniaxial stress. It must be taken into account, however, that the low $dT/dp$ values, along with a relatively large hysteresis, result in a poor reversibility of the barocaloric effect upon cycling, while good reversibility does indeed exist for the elastocaloric effect. With regards to the adiabatic temperature change, $\Delta T$ values are only moderate ($\Delta T \sim$ 4 K, inset in fig. \ref{figure1}), again reflecting the weak sensitivity of the martensitic transition to hydrostatic pressure.

\section{Concluding remarks}

Materials with giant mechanocaloric effects provide a good alternative for efficient and environmentally friendly cooling technologies and they complement the more mature magnetocaloric and electrocaloric ones. Shape memory alloys exhibit excellent mechanocaloric properties linked to their martensitic transformation. On the one hand, a number of magnetic shape memory alloys present $\Delta S$ and $\Delta T$ values that compare favourably to those reported for other giant barocaloric materials \cite{Manosa2011,Yuce2012,SternTaulats2014b,SternTaulats2015b,Matsunami2015,Lloveras2015}. On the other hand, the extreme sensitivity of the transition temperatures to uniaxial stress confers shape memory alloys unique elastocaloric properties.

In view of possible applications of shape memory alloys in cooling devices, the elastocaloric effect in these materials have the following important advantages in relation to magnetocaloric and electrocaloric materials. First, the strong sensitivity of the transition temperature to the applied stress enables the transition to be induced at temperatures far away from the stress-free transition temperature, which confer these alloys an extremely broad operation range which is limited at low values by the zero stress transition temperature and at higher temperatures by the limit of plasticity of the material. Values as large as 130 K have been reported for Cu-Zn-Al SMA which result in an outstanding refrigerant capacity of $\sim$ 2300 J/kg. Second, the elastocaloric effect in SMA exhibits an excellent cyclic reversibility: the large $\Delta S$ and $\Delta T$ values have been found to be very reproducible upon stress-cycling for a large number of cycles. It is worth emphasazing that ultralow fatigue Ti-Ni-Cu alloys have recently been developed for which the martensitic transition does not evolve for more than 10 million transformation  cycles \cite{Chluba2015}. 

While it is not foreseen that giant mechanocaloric materials will substitute electrocaloric and magnetocaloric ones, it is expected that each of them will have their own niche of applicability and the combination of all these phenomena should contribute to the gradual replacement of present refrigeration systems which feature lower efficiencies and are based on contaminant fluids.







\begin{acknowledgments} 
The authors acknowledge longstanding and fruitful collaboration with E. Stern-Taulats, E. Bonnot, R. Millan-Solsona, D. Soto-Parra, P.O. Castillo-Villa, R.Romero, X. Moya, E. Vives, T. Cast\'an, P. Lloveras, M. Barrio and  J.L. Tamarit.
\end{acknowledgments}





\begin{thebibliography}{9}

\bibitem{Fahler2012}  F\"ahler S.,R\"ossler UK., Kastner O., Eckert J., Eggeler G., Emmerich H., Entel P., M\"uller S., Quandt E., Karsten A. 2012. Caloric Effects in Ferroic Materials: New Concepts for Cooling. \textit{Adv. Eng. Mat.} \textbf{14}, 10.

\bibitem{Manosa2013a} Ma\~nosa L., Planes A., Acet M. 2013. Advanced Materials for solid-state refrigeration. \textit{J. Mater. Chem. A} \textbf{1}, 4925.

\bibitem{Moya2014} Moya X., Kar-Narayan S., Mathur ND. 2014. Caloric materials near ferroic phase transitions. \textit{Nature Mater.} \textbf{13}, 439.

\bibitem{Moya2015} Moya X., Defay E., Heine V., Mathur ND. 2015. Too cool to work. \textit{Nature Phys.} \textbf{11}, 202.

\bibitem{Kitanovski2015} Kitanovski A., Plaznik U., Tomc U., Poredos A. 2015. Present and future caloric refrigeration and heat-pump technologies. \textit{Int. J. Refrig.} \textbf{57}, 288.

\bibitem{Otsuka2002} Otsuka K., Kakeshita T. 2002. Science and Technology of Shape-Memory Alloys: New Develpoments. \textit{MRS Bull.} \textbf{27}, 91.

\bibitem{Bonnot2008} Bonnot E., Romero R., Ma\~nosa L., Vives E., Planes A. 2008. Elastocaloric Effect Associated with the Martensitic Transition in Shape-Memory Alloys. \textit{Phys. Rev. Lett.} \textbf{100}, 125901.

\bibitem{Manosa2008} Ma\~nosa L., Moya X., Planes A., Gutfleisch O., Lyubina J., Barrio M., Tamarit JL., Aksoy S., Krenke T., Acet M. 2008. Effects of hydrostatic pressure on the magnetism and martensitic transition of Ni-Mn-In magnetic superelastic alloys. 
\textit{Appl. Phys. Lett.}, \textbf{92}, 012515.

\bibitem{Manosa2010} Ma\~nosa L.,  Gonz\'alez-Alonso D., Planes A., Bonnot E., Barrio M., Tamarit JL., Aksoy S., Acet M. 2010. Giant solid-state barocaloric effect in the Ni-Mn-In magnetic shape-memory alloy. \textit{Nature Mater.} \textbf{9}, 478.

\bibitem{Planes2014} Planes A., Castan T., Saxena A. 2014. Thermodynamics of multicaloric effects in multiferroics. \textit{Philos. Mag.} \textbf{94}, 1893.

\bibitem{Casanova2005} Casanova F., Labarta A., Batlle X., P\'erez-Reche FJ., Vives E., Ma\~nosa L., Planes A. 2005. Direct observation of the magnetic-field-induced entropy change in Gd$_5$(Si$_x$Ge$_{1-x}$)$_4$ giant magnetocaloric alloys. \textit{Appl. Phys. Lett.} \textbf{86}, 262504.

\bibitem{SternTaulats2014a} Stern-Taulats E., Castillo-Villa PO., Ma\~nosa L., Frontera C., Pramanick S., Majumdar S., Planes A. 2014. Magnetocaloric effect in the low hysteresis Ni-Mn-In metamagnetic shape-memory alloy. \textit{J. Appl. Phys.} \textbf{115}, 173907.

\bibitem{Moya2013} Moya X., Stern-Taulats E., Crossley S., Gonz\'alez-Alonso D., Kar-Narayan S., Planes A., Ma\~nosa L., Mathur ND. 2013. Giant Electrocaloric Strength in Single-Crystal BaTiO$_3$. \textit{Adv. Mater.} \textbf{25}, 1360. 

\bibitem{Sandeman2012} Sandeman KG. 2012. Magnetocaloric materials: The search for new systems. \textit{Scripta Mater.} \textbf{67}, 566.

\bibitem{Otsuka1998} Otsuka K., Wayman CM. 1998. Shape Memory Materials. Cambridge University Press.

\bibitem{Planes2001} Planes A., Ma\~nosa L. 2001. Vibrational Properties of Shape-Memory Alloys. \textit{Solid State Phys.} \textbf{55}, 159.

\bibitem{Ahlers1986} Ahlers M. 1986. Martensite and equilibrium phases in Cu-Zn and Cu-Zn-Al alloys. \textit{Prog. Mat. Sci.} \textbf{30}, 135.

\bibitem{Manosa2009} Ma\~nosa L., Planes E., Vives E., Bonnot E., Romero R. 2009. The use of Shape-Memory Alloys for mechanical Refrigeration. \textit{Funct. Mater. Lett.} \textbf{2}, 73.

\bibitem{Vives2011} Vives E., Burrows S., Edwards RE., Dixon S., Ma\~nosa L., Planes A., Romero R. 2011. Temperature contour maps at the strain-induced martensitic transition of a Cu-Zn-Al shape-memory single crystal. \textit{Appl. Phys. Lett.} \textbf{98}, 011902.

\bibitem{Manosa2013b} Ma\~nosa L., Jarque-Farnos S., Vives E., Planes A. 2013. Large temperature span and giant refrigerant capacity in elastocaloric Cu-Zn-Al shape memory alloys. \textit{Appl. Phys. Lett.} \textbf{103}, 211904.

\bibitem{Tusek2015a} Tusek J., Engelbrecht K., Mill\'an-Solsona R., Ma\~nosa L., Vives E., Mikkelsen LP., Pryds N. 2015. The Elastocaloric Effect: A Way to Cool Efficiently. \textit{Adv. Energy Mater.} \textbf{5}, 1500361.

\bibitem{Obrado1997} Obrad\'o E., Ma\~nosa L., Planes A. 1997. Stability of the bcc phase of Cu-Al-Mn shape-memory alloys. \textit{Phys. Rev. B} \textbf{56}, 20.

\bibitem{Rodriguez1980} Rodr\'{\i}guez C., Brown LC. 1980. The Thermal Effect Due to Stress-Induced Martensite Formation in $\beta$-CuAlNi Single Crystals. \textit{Metall. Trans. A} \textbf{11A}, 147.

\bibitem{Vivesunp} Vives E., Ma\~nosa L., Planes A. unpublished.

\bibitem{Otsuka2005} Otsuka K.,  Ren X. 2005. Physical metallurgy of Ti-Ni-based shape memory alloys. \textit{Prog. Mater. Sci.} \textbf{50}, 511.

\bibitem{Frenzel2015} Frenzel J., Wieczorek A., Ophale I., Mass B., Drautz R., Eggeler G. 2015. On the effect of alloy composition on martensite start temperatures and latent heats in Ni-Ti-based shape memory alloys. \textit{Acta Mater.} \textbf{90}, 213.

\bibitem{SotoParra2016} Soto-Parra D., Vives E., Ma\~nosa L., Matutes-Aquino JA., Flores-Z\'u\~niga H., Planes A. 2016. Elastocaloric effect in Ti-Ni shape-memory wires associated with the B2-B19' and B2-R structural transitions. \textit{Appl. Phys. Lett.} \textbf{108}, 071902.

\bibitem{Bechtold2012} Bechtold C., Chluba C., Lima de Miranda R., Quandt E. 2012. High cyclic stability of the elastocaloric effect in sputtered TiNiCu shape memory films. \textit{Appl. Phys. Lett.} \textbf{101}, 091903.

\bibitem{Cui2012} Cui J., Wu Y., Muehlbauer J., Hwang Y., Radermacher R., Fackler S., Wuttig M., Takeuchi I. 2012. Demonstration of high efficiency elastocaloric cooling with large $\Delta$T using NiTi wires.
\textit{Appl. Phys. Lett.} \textbf{101}, 073904.

\bibitem{Pieczyska2013} Pieczyska EA., Tobushi H., Kulasinski K. 2013. Development of transformation bands in TiNi SMA for various stress and strain rates studied by a fast and sensitive infrared camera. \textit{Smart Mater. Struct.} \textbf{22}, 035007.

\bibitem{Tusek2015b} Tusek J., Engelbrecht K., Mikkelsen LP., Pryds N. 2015. Elastocaloric effect of Ni-Ti wire for application in a cooling device. \textit{J. Appl. Phys.} \textbf{117}, 124901.

\bibitem{Ossmer2015b} Ossmer H., Miyazaki S., Kohl M. 2015. The elastocaloric effect in TiNi-based foils. \textit{Materials Today: Proc.} \textbf{2S}, S971.

\bibitem{Ossmer2014} Ossmer H., Lambrecht F.,  Gueltig M., Chluba C., Quandt E., Kohl M. 2014.  Evolution of the temperature profiles in TiNi films for elastocaloric cooling. \textit{Acta Mater.} \textbf{81}, 9.

\bibitem{Ossmer2015a} Ossmer H., Chluba C., Gueltig M., Quandt E., Kohl M. 2015. Local Evolution of the Elastocaloric Effect in TiNi-Based Films. \textit{Shap. Mem. Superelasticity} \textbf{1}, 142.

\bibitem{Schmidt2015} Schmidt M., Ullrich J., Wieczorek A., Frenzel J., Sch\"utze A., Eggeler G., Seelecke S. 2015. Thermal Stabilization of NiTiCuV Shape Memory Alloys: Observations During Elastocaloric Training. \textit{Shap. Mem. Superelasticity} \textbf{1}, 132.

\bibitem{Pataky2015} Pataky GJ., Ertekin E., Sehitoglu H. 2015. Elastocaloric cooling potential of NiTi, Ni$_2$FeGa and CoNiAl. \textit{Acta Mater.} \textbf{96}, 420.

\bibitem{Tang2015} Tang Z., Wang Y., Liao X., Wang D., Yang S., Song X. 2015. Stress dependent transforming behaviours and associated functional properties of a nano-precipitates induced strain glass alloy. \textit{J. Alloys and Comp.} \textbf{622}, 622.

\bibitem{Kakeshita2002} Kakeshita T., Ullako K. 2002. Giant Magnetostriction in Ferromagnetic Shape-Memory Alloys. \textit{MRS Bull.} \textbf{27}, p. 105.

\bibitem{Ullakko1996} Ullakko K., Huang JK., Knatner C., O'Handley RC., Kokorin VV. 1996. Large magnetic-field-induced strains in Ni$_2$MnGa single crystals. \textit{Appl. Phys. Lett.} \textbf{69}, 1966

\bibitem{Planes2009} Planes A., Ma\~nosa L., Acet M. 2009. Magnetocaloric effect and its relation to shape-memory properties in ferromagnetic Heusler alloys. \textit{J. Phys. Condens. Matter}, \textbf{21}, 233201.

\bibitem{Acet2011} Acet M., Ma\~nosa L., Planes A. 2011. Magnetic-Field-Induced Effects in Martensitic Heusler-based Magnetic Shaper Memory Alloys. \textit{Handbook of Magnetic Materials}, \textbf{19}, 231.

\bibitem{Khovaylo2009} Khovaylo VV., Kanomata T., Tanaka T., Nakashima M., Amako Y., Kainuma R., Umetsu RY., Morito H., Miki H. 2009. Magnetic properties of Ni$_{50}$Mn$_{34.8}$In$_{15.2}$ probed by M\"ossbauer spectroscopy. \textit{Phys. Rev. B} \textbf{80}, 144409.

\bibitem{Ollefs2015} Ollefs K., Sch\"oppner Ch., Titov I., Meckenstock R., Wilhelm F., Rogalev A., Liu J., Cutfleisch O., Farle M., Wende H., Acet A. 2015. Magnetic ordering in magnetic shape memory alloy Ni-Mn-In-Co. \textit{Phys. Rev. B} \textbf{92}, 224429.

\bibitem{Aksoy2009} Aksoy A., Acet M., Deen PP., Ma\~nosa A., Planes A. 2009. Magnetic correlations in martensitic Ni-Mn-based Heusler shape-memory alloys: Neutron polarization analysis. \textit{Phys. Rev. B} \textbf{79}, 212401.

\bibitem{Kainuma2006} Kainuma R., Imano Y., Ito W., Sutou Y., Morito H., Okamoto S., Kitakami O., Oikawa K., Fujita A., Kanomata T., Ishida K. 2006. Magnetic-field-induced shape recovery by reverse phase transformation. \textit{Nature}, \textbf{439}, 957

\bibitem{Krenke2007} Krenke T., Duman E., Acet M., Wassermann EF., Moya X., Ma\~nosa L., Planes A., Suard E., Ouladdiaf B. 2007. Magnetic superelasticity and inverse magnetocaloric effect in Ni-Mn-In. \textit{Phys. Rev. B} \textbf{75}, 104414.

\bibitem{Krenke2005} Krenke T., Duman E., Acet M., Wassermann EF., Moya X., Ma\~nosa L., Planes A. 2005. Inverse magnetocaloric effect in ferromagnetic Ni-Mn-Sn alloys. \textit{Nature Mater.} \textbf{4}, 450.

\bibitem{Cui2004} Cui J., Shield TW., James RD. 2004. Phase transformation and magnetic anisotropy of an iron-palladium ferromagnetic shape-memory alloy. \textit{Acta Mater.}, \textbf{52}, 35.

\bibitem{Aksoy2007} Aksoy S., Krenke T., Acet M., Wassermann EF., Moya X., Ma\~nosa L., Planes A. 2007. Magnetization easy axis in martensitic Heusler alloys estimated by strain measurements under  magnetic field. \textit{Appl. Phys. Lett.}, \textbf{91}, 251915.

\bibitem{Kihara2014} Kihara T., Xu X., Ito W., Kainuma R., Tokunaga M. 2014. Direct measurements of inverse magnetocaloric effects in metamagnetic shape-memory alloy NiCoMnIn. \textit{Phys. Rev. B} \textbf{90}, 214409.

\bibitem{CastilloVilla2011} Catillo-Villa PO., Soto-Parra DE., Matutes-Aquino JA., Ochoa-Gamboa RA., Planes A., Ma\~nosa L., Gonz\'alez-Alonso D., Stipcich M., Romero R., R\'{\i}os-Jara D., Flores-Z\'u\~niga H. 2011. Caloric effects induced by magnetic and mechanical fields in a Ni$_{50}$Mn$_{25-x}$Ga$_{25}$Co$_x$ magnetic shape memory alloy. \textit{Phys. Rev. B} \textbf{83}, 174109.

\bibitem{SotoParra2010} Soto-Parra DE., Vives E., Gonz\'alez-Alonso D., Ma\~nosa L., Planes A., Romero R.,  Matutes-Aquino JA., Ochoa-Gamboa RA., Flores-Z\'u\~niga H. 2010. Stress- and magnetic field-induced entropy changes in Fe-doped Ni-Mn-Ga shape-memory alloys. \textit{Appl. Phys. Lett.} \textbf{96}, 071912.

\bibitem{Xu2015} Xu Y., Lu B., Sun W., Yan A., Liu J., 2015. Large and reversible elastocaloric effect in dual-phase Ni$_{54}$Fe$_{19}$Ga$_{27}$ superelastic alloys. \textit{Appl. Phys. Lett.} \textbf{106}, 201903.

\bibitem{Xiao2015b} Xiao F., Jin M., Liu J., Jin X., 2015. Elastocaloric effect in Ni$_{50}$Fe$_{19}$Ga$_{27}$Co$_{4}$ single crystals. \textit{Acta Mater.} \textbf{96}, 292.

\bibitem{Huang2015} Huang YJ., Hu HD., Bruno NM., Chen JH., Karaman I., Ross JH., Li JG. 2015. Giant elastocaloric effect in directionally solidified Ni-Mn-In magnetic shape memory alloy. \textit{Scripta Mater.} \textbf{105}, 42.

\bibitem{Lu2014} Lu B., Xiao F., Yan A., Liu J. 2014. Elastocaloric effect in a textured polycrystalline Ni-Mn-In-Co metamagnetic shape memory alloy. \textit{Appl. Phys. Lett.} \textbf{105}, 161905.

\bibitem{Lu2015} Lu B., Zhang P., Xu Y., Sun W., Liu J. 2015. Elastocaloric effect in Ni$_{45}$Mn$_{36.4}$In$_{13.6}$Co$_5$ metamagnetic shape memory alloys under mechanical cycling. \textit{Mater. Lett.} \textbf{148}, 110.

\bibitem{MillanSolsona2014} Mill\'an-Solsona R., Stern-Taulats E., Vives E., Planes A., Sharma J., Nayak AK., Suresh KG., Ma\~nosa L. 2014. Large entropy change associated with the elastocaloric effect in polycrystalline Ni-Mn-Sb-Co magnetic shape memory alloys. \textit{Appl. Phys. Lett.} \textbf{105}, 241901.

\bibitem{Sun2016} Sun W., Liu J., Lu B., Li Y., Yan A. 2016. Large elastocaloric effect at small transformation strain in Ni$_{45}$Mn$_{44}$Sn$_{11}$ metamagnetic shape memory alloys. \textit{Scripta Metall.} \textbf{114}, 1.

\bibitem{CastilloVilla2013} Castillo-Villa PO., Ma\~nosa L., Planes A., Soto-Parra DE., S\'anchez-Llamazares JL., Flores-Zu\~niga H., Frontera C. 2013. Elastocaloric and magnetocaloric effects in Ni-Mn-Sn(Cu) shape-memory alloy. \textit{J. Appl. Phys. } \textbf{113}, 053506.

\bibitem{Xiao2013} Xiao F., Fukuda T., Kakeshita T. 2013. Significant elastocaloric effect in a Fe-31.2Pd (at \%) single crystal. \textit{Appl. Phys. Lett.} \textbf{102}, 161914.

\bibitem{Xiao2015a} Xiao F., Fukuda T., Kakeshita T., Jin X. 2015. Elastocaloric effect by a weak first-order transformation associated with lattice softening in an Fe-31.2Pd (at \%) alloy. \textit{Acta Mater.} \textbf{87}, 8.

\bibitem{Krenke2005b}  Krenke T., Acet M., Wassermann EF., Moya X., Ma\~nosa L., Planes A. 2005. Martensitic transitions and the nature of ferromagnetism in the austenitic and martensitic states of Ni-Mn-Sn alloys. \textit{Phys. Rev. B} \textbf{72}, 014412.

\bibitem{Krenke2006} Krenke T., Acet M., Wassermann EF., Moya X., Ma\~nosa L., Planes A. 2006. Ferromagnetism in the austenitic and martensitic states of Ni-Mn-In alloys. \textit{Phys. Rev. B} \textbf{73}, 174413.

\bibitem{Xiao2015c} Xiao F., Fukuda T., Kakeshita T. 2015. Critical point of martensitic transformation under stress in an Fe-31.3Pd (at \%) shape memory alloy. \textit{Philos. Mag.} \textbf{95}, 1390.

\bibitem{SternTaulats2015a} Stern-Taulats E., Planes A., Lloveras P., Barrio M., Tamarit JL., Pramanick S., Majumdar S., Y\"uce S., Emre B., Frontera C. 2015. Tailoring barocaloric and magnetocaloric properties in low-hysteresis magnetic shape memory alloys. \textit{Acta Mater.} \textbf{96}, 324.

\bibitem{Manosa2014} Ma\~nosa L., Stern-Taulats E., Planes A., Lloveras P., Barrio M., Tamarit JL., Emre B., Y\"uce S., Fabbrici S., Albertini F. 2014. Barocaloric effect in metamagnetic shape memory alloys. \textit{Phys. Stat. Sol. B} \textbf{251}, 2114.


\bibitem{Manosa2011} Ma\~nosa L., Gonz\'alez-Alonso D., Planes A., Barrio M., Tamarit JL., Titov IS., Acet M., Bhattacharyya A., Majumdar S. 2011. Inverse barocaloric effect in the giant magnetocaloric La-Fe-Si-Co compound. \textit{Nature Comm.} \textbf{2}, 595.

\bibitem{Albertini2007} Albertini F., Kamar\'ad J., Arnold Z., Pareti L., Villa E., Righi L. 2007. Pressure effects on the magnetocaloric properties of Ni-rich and Mn-rich Ni$_2$MnGa alloys. \textit{J. Magn. Mag. Mat.} \textbf{316}, 364.

\bibitem{Mandal2009} Mandal K., Pal D., Scheerbaum N., Lyubina J., Gutfleisch O. 2009. Effect of pressure on the magnetocaloric properties of nickel-rich Ni-Mn-Ga Heusler alloys. \textit{J. Appl. Phys.} \textbf{105}, 073509.

\bibitem{Kamarad2005} Kamar\'ad J., Albertini F., Arnold Z., Casoli F., Pareti L., Paoluzi A. 2005. Effect of hydrosttic pressure on magnetization of Ni$_{2+x}$Mn$_{1-x}$Ga alloys. \textit{J. Magn. Mag. Mat.} \textbf{290-291}, 669.

\bibitem{Yasuda2007} Yasuda T., Kanomata T., Saito T., Yosida H., Nishihara H., Kainuma R., Oikawa K. Ishida K., Neumann KU., Ziebeck KRA. 2007. Pressure effect on transformation temperatures of ferromagnetic shape memory alloy Ni$_{50}$Mn$_{36}$Sn$_{14}$. \textit{J. Magn. Mag. Mat.} \textbf{310}, 2770.

\bibitem{Muthu2011} Esakki Muthu S., Rama Rao NV., Manivel Raja M., Arumugam S., Matsubayasi K., Uwatoko Y., 2011. Hydrostatic pressure effect on the martensitic transition, magnetic, and magnetocaloric properties in  Ni$_{50-x}$Mn$_{37+x}$Sn$_{13}$ Heusler alloys. \textit{J. Appl. Phys.} \textbf{110}, 083902.

\bibitem{Sharma2011} Sharma VK., Chattopadhyay MK., Roy SB. 2011. The effect of external pressure on the magnetocaloric effect of Ni-Mn-In alloy. \textit{J. Phys.: Condens. Matter} \textbf{23}, 366001.

\bibitem{Yuce2012} Y\"uce S., Barrio M., Emre B., Stern-Taulats E., Planes A., Tamarit JL., Mudryk Y., Gschneidner KA., Pecharsky VK., Ma\~nosa L. 2012. Barocaloric effect in magnetocaloric prototype Gd$_5$Si$_2$Ge$_2$. \textit{Appl. Phys. Lett.} \textbf{101}, 071906.


\bibitem{SternTaulats2014b} Stern-Taulats E., Planes A., Lloveras P., Barrio M., Tamarit JL., Pramanick S., Majumdar S., Frontera C., Ma\~nosa L. 2014. Barocaloric and magnetocaloric effects in Fe$_{49}$Rh$_{51}$. \textit{Phys. Rev. B} \textbf{89}, 214105.

\bibitem{SternTaulats2015b} Stern-Taulats E., Gr\`acia-Condal A., Planes A., Lloveras P., Barrio M., Tamarit JL., Pramanick S., Majumdar S., Ma\~nosa L. 2015. Reversible adiabatic temperature changes at the magnetocaloric and barocaloric effects in Fe$_{49}$Rh$_{51}$. \textit{Appl. Phys. Lett.} \textbf{107}, 152409.

\bibitem{Matsunami2015} Matsunami D., Fujita A., Takenaka K., Kano M. 2015. Giant barocaloric effect enhanced by the frustration of the antiferromagnetic phase in Mn$_3$GaN. \textit{Nature Mater.} \textbf{14}, 73.

\bibitem{Lloveras2015} Lloveras P., Stern-Taulats E., Barrio M., Tamarit JL., Crossley S., Li W., Pomjakushin V., Planes A., Ma\~nosa L., Mathur ND., Moya X. 2015. Giant barocaloric effects at low pressure in ferrielectric ammonium sulphate. \textit{Nature Comm.} \textit{6}, 8801.

\bibitem{Chluba2015} Chluba C., Ge W., Lima de Miranda R., Strobel J., Kienle L., Quandt E., Wuttig M. 2015. Ultralow-fatigue shape memory alloy films. \textit{Science} \textbf{348}, 1004.

\end{thebibliography}
\end{document}